\documentclass[aps,pre,twocolumn,showpacs,superscriptaddress]{revtex4}
\usepackage{epsfig}
\usepackage{times}
\begin{document}

\title{Risk-driven migration and the collective-risk social dilemma}

\author{Xiaojie Chen}
\email{chenx@iiasa.ac.at}
\affiliation{Evolution and Ecology Program, International Institute for Applied Systems Analysis (IIASA), Schlossplatz 1, A-2361 Laxenburg, Austria}
\author{Attila Szolnoki}
\email{szolnoki.attila@ttk.mta.hu}
\affiliation{Institute of Technical Physics and Materials Science, Research Centre for Natural Sciences, Hungarian Academy of Sciences, P.O. Box 49, H-1525 Budapest, Hungary}
\author{Matja\v{z} Perc}
\email{matjaz.perc@gmail.com}
\affiliation{Faculty of Natural Sciences and Mathematics, University of Maribor, Koro{\v s}ka cesta 160, SI-2000 Maribor, Slovenia}

\begin{abstract}
A collective-risk social dilemma implies that personal endowments will be lost if contributions to the common pool within a group are too small. Failure to reach the collective target thus has dire consequences for all group members, independently of their strategies. Wanting to move away from unfavorable locations is therefore all but surprising. Inspired by these observations, we here propose and study a collective-risk social dilemma where players are allowed to move if the collective failure becomes too probable. More precisely, this so-called risk-driven migration is launched depending on the difference between the actual contributions and the declared target. Mobility therefore becomes an inherent property that is utilized in an entirely self-organizing manner. We show that under these assumptions cooperation is promoted much more effectively than under the action of manually determined migration rates. For the latter, we in fact identify parameter regions where the evolution of cooperation is incredibly inhibited. Moreover, we find unexpected spatial patterns where cooperators that do not form compact clusters outperform those that do, and where defectors are able to utilize strikingly different ways of invasion. The presented results support the recently revealed importance of percolation for the successful evolution of public cooperation, while at the same time revealing surprisingly simple ways of self-organization towards socially desirable states.
\end{abstract}

\pacs{89.75.Fb, 87.23.Ge, 87.23.Kg}

\maketitle

\section{Introduction}
Our remarkable other-regarding abilities are believed to have evolved because alone we would have difficulties rearing offspring that survived \cite{hrdy_11}. Between-group conflicts are also cited frequently to that effect \cite{fu_srep12}. Although today the main challenges appear to be different, cooperation and in-group solidarity are still very much desired and indeed urgently needed behavioral traits \cite{nowak_11}. Were it not for them, the preservation of natural resources for future generations, the provisioning of health and social care, as well as many other amenities we have come to take for granted, would be greatly challenged. The ``tragedy of the commons'' \cite{hardin_g_s68} can only be averted if we manage to withstand the temptations of free-riding on the efforts of others, and if we succeed in fully realizing the global long-term implications of our shortsighted behavior that is aimed only at maximizing our current wellbeing \cite{garas_njp10}. History teaches us that this is a formidable ask, mainly because it goes against our natural instincts.

Theoretical insights on the subject are to a large extent due to evolutionary game theory \cite{hofbauer_98, nowak_06}, which has proven times and again as a very competent framework for the study of the evolution of cooperation \cite{axelrod_84}. Different branches of science, ranging from biology, sociology and economics to mathematics and physics, converge on this particular problem, as evidenced by reviews that comprehensively describe recent advances \cite{doebeli_el05, nowak_s06, szabo_pr07, schuster_jbp08, roca_plr09, perc_bs10}. The public goods game, in particular, is frequently considered as the paradigm that succinctly captures the essential social dilemma that emerges as a consequence of group and individual interests being inherently different \cite{sigmund_10}. Governed by group interactions, the public goods game requires that players decide simultaneously whether they wish to contribute to the common pool or not. Regardless of the chosen strategy, each member of the group receives an equal share of the public good after the initial investments are multiplied by a synergy factor that takes into account the added value of collaborative efforts. In generally framed in coevolutionary models \cite{perc_bs10,lozano_acs12}, recent research on the spatial public goods game \cite{szolnoki_pre09c} highlighted appropriate partner selection \cite{wu_t_pre09, zhang_hf_epl11}, diversity \cite{santos_n08, yang_hx_pre09, ohdaira_acs11, santos_jtb12}, the critical mass \cite{szolnoki_pre10}, heterogeneous wealth distributions \cite{wang_j_pre10b}, the introduction of punishment \cite{brandt_prsb03, helbing_ploscb10, hilbe_srep12} and reward \cite{szolnoki_epl10}, network modularity \cite{gomez-gardenes_c11}, coordinated investments \cite{vukov_jtb11} and conditional strategies \cite{szolnoki_pre12}, or the joker effect \cite{arenas_jtb11,requejo_pre12} as viable to promote the evolution of public cooperation.

Although deceptively similar to the public goods game, the collective-risk social dilemma captures more vividly an additional important feature of ``problems of the commons'' \cite{milinski_pnas08}, which is that failure to reach a declared global target can have severe long-term consequences. Opting out of carbon emission reduction to harvest short-term economic benefits is a typical example. The blueprint of the game is as follows. All players are considered to have an initial endowment, and cooperation means contributing a fraction of it to the common pool. Defectors do not contribute. The risk level is determined by a collective target that should be reached with individual investments. If a group fails to reach this target, all members of the group loose their remaining endowments with a certain probability, while otherwise the endowments are retained. Most importantly, the risk of collective failures has been identified as a potent promoter of responsible social behavior \cite{milinski_pnas08, wang_j_pre09, santos_pnas11, greenwood_epl11, raihani_cc11}. Here we employ the collective-risk social dilemma, where the risk is a dynamically changing group-performance-dependent quantity, which can be fine-tuned by a single parameter $\beta$ that interpolates between a step-like and a flat risk function.

Considering structured populations, as pioneered by Nowak and May \cite{nowak_n92b}, also invites mobility as an inherent property of players that adds another layer of reality to the study. Clearly attesting to this fact is the overwhelming attention mobility has received, both in games governed by pairwise \cite{kerr_n06,vainstein_jtb07, wu_zx_pre09, sicardi_jtb09, droz_epjb09, helbing_epjb09,helbing_pnas09, meloni_pre09, cheng_hy_njp10,jiang_ll_pre10, yang_hx_pre10, yu_w_pre11, wu_t_pone11, cheng_hy_njp11, lin_yt_pa11} as well as, although to a much lesser extent, in games governed by group interactions \cite{roca_pnas11, wu_t_pre12, cardillo_pre12}. Indeed, being free and mobile is one of the hallmarks of modern societies. However, family and friend ties, traveling and adaptation costs, as well as all the other difficulties that can be associated with migration act as strong inhibitors of changing location. Here we take this into account by introducing and studying so-called risk-driven migration. The main premise lies in the assumption that we are \textit{forced} to migrate by the immediate (possibly unfavorable) environment rather than this being the consequence of our explicit desire to do so. As noted, the risk depends on the difference between the actual contributions and the declared target in each group, which changes dynamically as do the spatial patterns. Players are aware of this risk, and they are more likely to move the higher the risk. This is significantly different from previous studies, where for example success-driven migration \cite{helbing_pnas09} or identical migration rates for all players were assumed \cite{vainstein_jtb07, sicardi_jtb09}. In our case migration is a self-organizing process that depends only on the risk each individual is exposed to. Thus, unlike in previous game-theoretical models, it does not require additional parameters, nor does it assume additional cognitive skills that would be needed by players to estimate the income at potential new locations. As we will show in what follows, this self-organizing risk-driven migration outperforms previous migratory actions, and it leads to spatial patterns that enable cooperators to optimally exploit their mutually beneficial behavior. Before presenting the main results, however, we proceed with the details of the model in the next section.

\section{Model}
We consider the collective-risk social dilemma on a square lattice of size $L\times L$ with periodic boundary conditions. Each site can either be empty or occupied by a cooperator or defector. The fraction of occupied sites constitutes the population density $\rho$ ($0<\rho\leq 1$), which is kept constant during the evolutionary process. Initially thus a joint total of $\rho L^2$ cooperators and defectors populate the lattice uniformly at random with equal probability, and each of them is granted an initial endowment $b$. Without losing generality, we use $b=1$ throughout this work.

We employ asynchronous updating, such that a randomly selected player $x$ plays the collective-risk social dilemma game with its eight nearest neighbors (if present), and thereby collects its total payoff $P_x$. Cooperators contribute an amount $c<b$ from the endowment to the public good, whereas defectors contribute nothing. Due to the vacant sites there is an obvious possibility for a different number of players participating in each collective-risk social dilemma. The collective target for each group is therefore defined with this in mind, such that $T=n\alpha$, where $n$ is the number of active players and $\alpha$ $(0\leq \alpha \leq 1)$ is the weighting factor that determines the collective threshold. If the total amount of collected contributions within a group $G_i$ is either reached or surpassed, each player can keep what it has not yet contributed into the common pool. In the opposite case, if the collective target is not reached, all group members loose their remaining endowments with a probability $r_i$. The corresponding risk probability in a group is calculated according to
\begin{equation} r_i=\left\{
\begin{array}{lll}
(\frac{T-n_c}{T})^{\beta} & \mbox{ if }n_c<T,\\
0 & \mbox{ if }n_c\geq T,
\end{array} \right.
\end{equation}
where $n_c$ is the number of cooperators in the group, while $\beta$ is a tunable parameter determining the nonlinearity of the risk function. For $\beta=0$ the risk function is step-like,
which was most frequently considered in previous works. For $\beta=1$, on the other hand, the
risk function is linearly proportional to the difference between the collective target and the collected contributions. Evidently, $\beta$ can also be larger than one, although in this case the collective risk is so small that the evolution of cooperation is strongly inhibited. We therefore focus on the unit interval of $\beta$.

Following the accumulation of the payoff, player $x$ moves to a randomly chosen empty site (if it exists) within its Moore neighborhood with the probability $r_m=\Sigma_i r_i/n$, which simply quantifies the average risk the player experiences by being in its current location. If there are no empty sites within the Moore neighborhood player $x$ does not move. On the other hand, to avoid long periods of isolation, players with no neighbors must make a mandatory move, although this is a technical detail that does not notably influence the outcome of the game. At this point we emphasize again that this simple definition of mobility introduces no additional parameters, it does not assume special cognitive skills of the players, and it cannot be subject to different mobility time-scale investigations. Interaction of time scales are known to play an important role in evolutionary dynamics \cite{pacheco_prl06, roca_prl06, szolnoki_epjb09}, yet in our case the propensity to move can vary significantly from player to player, and of course it varies also over time. It is in fact a self-organizing rather than a manually adjusted process, and as we will show, performs superior even under the most adverse conditions.

\begin{figure}
\centering
\includegraphics[width=8cm]{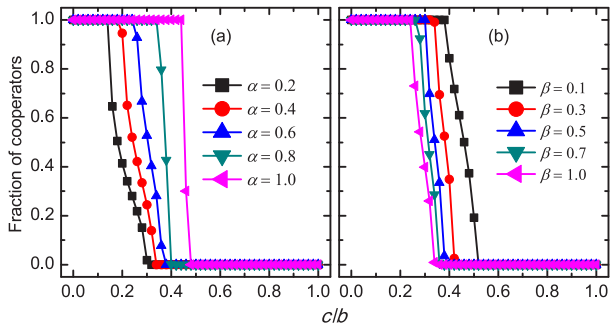}
\caption{(Color online) Fraction of cooperators as a function of the donation ratio $c/b$ for a fixed value of $\beta=0.5$ and different values of $\alpha$ in (a), and for a fixed value of $\alpha=0.7$ and different values of $\beta$ in (b). In both panels the density of players is $\rho=0.5$.}
\label{fig1}
\end{figure}

\begin{figure*}
\centering
\includegraphics[width=16.5cm]{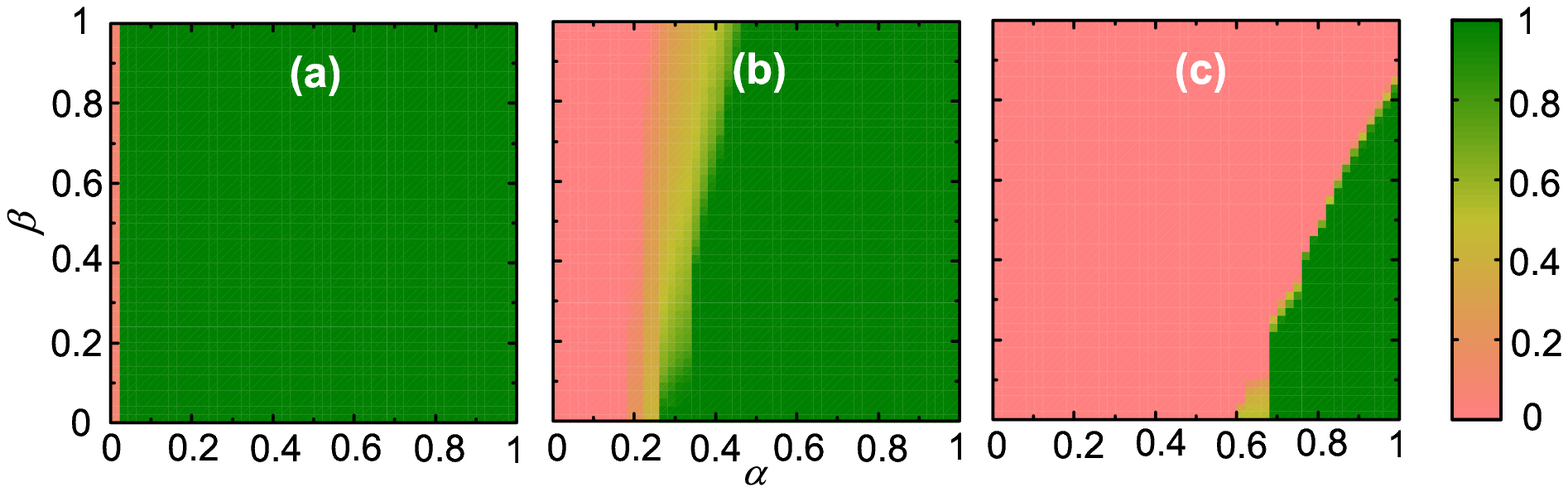}
\caption{(Color online) Fraction of cooperators in dependence on $\alpha$ and $\beta$ in a contour plot form for three different donation ratios: (a) $c/b=0.1$, (b) $c/b=0.3$ and (c) $c/b=0.5$. In all three panels the density of players is $\rho=0.3$.}
\label{fig2}
\end{figure*}

Lastly then, player $x$ adopts the strategy of a randomly chosen neighbor $y$ with a probability \begin{equation}
f(P_y-P_x)=\frac{1}{1+\exp[-(P_y-P_x)/\kappa]},
\end{equation}
where $\kappa$ denotes the amplitude of noise \cite{szabo_pr07}. Taking into account the insight of \cite{szolnoki_pre09c}, we set $\kappa=0.5$, which implies that better performing strategies will very likely spread, yet it is not impossible for a strategy performing worse to spread either. It is worth pointing out that the results reported below remain qualitative identical if the best-takes-over rule \cite{nowak_n92b} is used for strategy updating.

Results reported in the next section have been obtained by means of Monte Carlo simulations, which were carried out using $L=50$ and $100$ different initial conditions (the same results can of course also be obtained if using larger lattices). As is standard practice, during one full Monte Carlo step, all individuals will have received a chance once on average to learn a new strategy from one of their neighbors.

\section{Results}

Figure~\ref{fig1}(a) shows the fraction of cooperators at equilibrium as a function of the donation ratio $c/b$ at $\beta=0.5$ and $\rho=0.5$ for five different values of $\alpha$. It can be observed that for each value of $\alpha$ the fraction of cooperators decreases with increasing $c/b$. In addition, for some intermediate $c/b$ values the fraction of cooperators increases with increasing
$\alpha$. In Fig.~\ref{fig1}(b) we show the fraction of cooperators as a function of the donation ratio $c/b$ for a fixed value of $\alpha=0.7$ and five different values of $\beta$. It can be observed that for each value of $\beta$ the fraction of cooperators decreases with increasing
$c/b$. At some intermediate $c/b$ values, however, the fraction of cooperators decreases with increasing $\beta$.

In order to explore the effects of $\alpha$ and $\beta$ more precisely, we show in Fig.~\ref{fig2} the fraction of cooperators in dependence on $\alpha$ and $\beta$ together for three different values of $c/b$. We see that for small $c/b$ (for example $c/b=0.1$) the fraction of cooperators is zero when $\alpha$ is zero, regardless of the value of $\beta$. In fact, in this case there is no collective-risk, and our model is identical to the traditional public goods game. Correspondingly, defectors can have a higher payoff as long as $c/b>0$, and thus cooperators cannot survive. However, in other parameter regions full cooperation can be achieved, even for small $\alpha>0$ or large $\beta$ [Fig.~\ref{fig2}(a)]. When the donation ratio becomes larger (for example $c/b=0.3$), cooperators cannot survive for small $\alpha$ [Fig.~\ref{fig2}(b)]. However, full cooperation can still be achieved for large $\alpha$, although the region shrinks. As $c/b$ continues to increase (for example $c/b=0.5$), cooperators can only survive for large $\alpha$ and small $\beta$. For still larger $c/b$, full defection is obtained irrespective of $\alpha$ and $\beta$ (not shown here).

From this it follows that, since for small $\alpha$ the risk of collective failure is small, cooperators do not have a higher payoff than the neighboring defectors, especially for higher $c/b$. In addition, individuals are easily satisfied with their current positions, and accordingly they do not move. In this situation, cooperators are easily exploited by defectors and are indeed wiped out relatively fast. But as $\alpha$ increases, the risk of collective failure is increases as well. Consequently, there is thus a higher probability that cooperators will have a similar payoff as the surrounding defectors. At the same time, the migration becomes much more probable, so that interactions between cooperators and defectors become less frequent. The effectiveness at which defectors exploit cooperators therefore becomes smaller, and accordingly, the successful evolution of public cooperation is likelier.

When $\alpha$ is larger, the randomly selected defectors are more frequently dissatisfied with their current sites than their neighboring cooperators, especially when they encounter other defectors. Thus, clusters of defectors easily collapse in this model because a higher fraction of them changes their position during one Monte Carlo step. However, when $\beta$ is small, a tiny fraction of selected cooperators will move at the beginning of the evolutionary process. These cooperators are therefore able form small yet supporting clusters after the defectors depart, and they can in fact resist the invasion. More to the point, cooperators can use the safe haven of aggregated clusters to invade mobile defectors, ultimately resulting in an increasing density. In agreement with this argument defectors begin moving even more frequently, yet also less steadily. In the final stages of the evolutionary process migrations seize and the cooperators begin dominating the whole population. When $\beta$ becomes larger, a high fraction of randomly selected defectors still changes their locations, and quite surprisingly some of the selected cooperators also move to nearby empty sites. This indicates that, under such circumstances, cooperators cannot form effective clusters as easily as they have done before. In fact, even if some clusters succeed in forming, the mobile defectors can destroy them effectively. As we will elaborate later, such conditions enable a dynamical equilibrium between cooperators and defectors.

\begin{figure*}
\centering
\includegraphics[width=16cm]{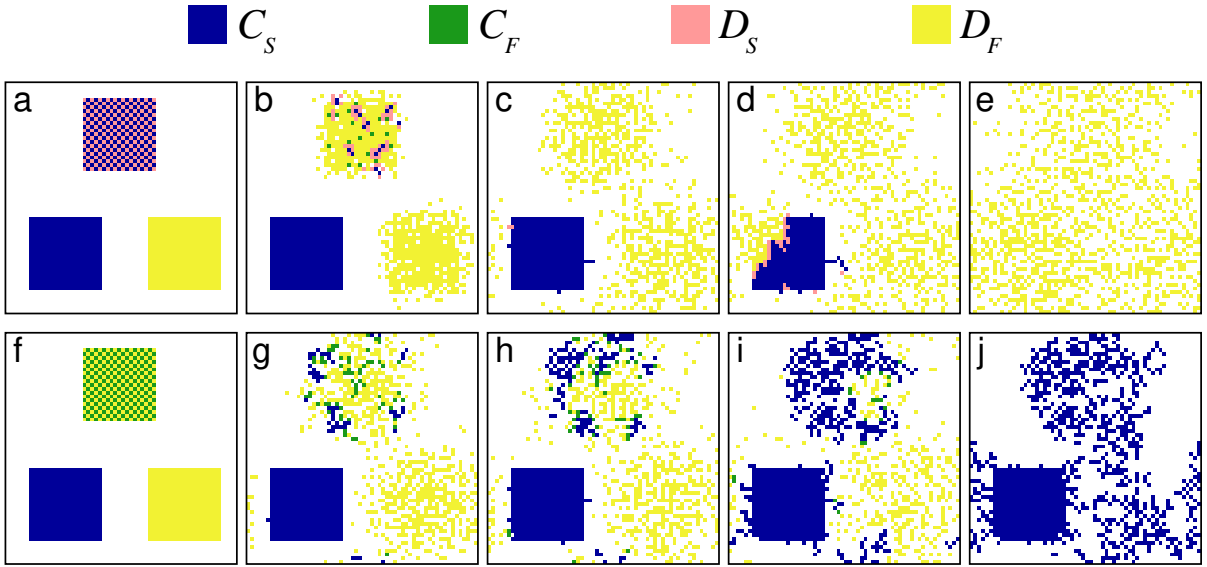}
\caption{(Color online) Pattern formation as observed from a prepared initial state at $\beta=0.2$. The weighting factor related to the collective threshold is $\alpha=0.2$ [upper row from (a) to (e)] and $\alpha=0.9$ [bottom row from (f) to (j)]. Cooperators whose focal group succeeds (fails) to reach the collective target are denoted blue (green). Similarly, defectors whose focal group succeeds (fails) are denoted pink (yellow). For easier referencing, the color legend is provided also on the top of the figure. White are empty sites. Other parameters are: $\rho=0.30$, $c/b=0.5$ and $L=60$. Monte Carlo steps are $t=0, 3, 25, 40, 90$ for panels (a) to (e), and $t=0, 18, 30, 53, 465$ for panels (f) to (j), respectively.}
\label{fig3}
\end{figure*}

To demonstrate the leading mechanisms that determine pattern formation at different $\alpha - \beta$ pairs, we present representative sequences of snapshots, as obtained from prepared initial states depicted in Figs.~\ref{fig3}(a) and (f). Here, as a starting configuration, we begin with three domains in which the strategies are either pure $C$ or $D$ (the two bottom islands) or the mixture of these two strategies (the single upper island). To clarify the importance of the collective risk, we have applied two different colors for each strategy in order to mark the success of the group surrounding the focal player. More precisely, we mark by blue (green) a cooperator whose central group succeeds (fails) to reach the collective target. Similarly, pink (yellow) are defectors in the center of a successful low risk (unsuccessful high risk) group.

\begin{figure*}
\centering
\includegraphics[width=16cm]{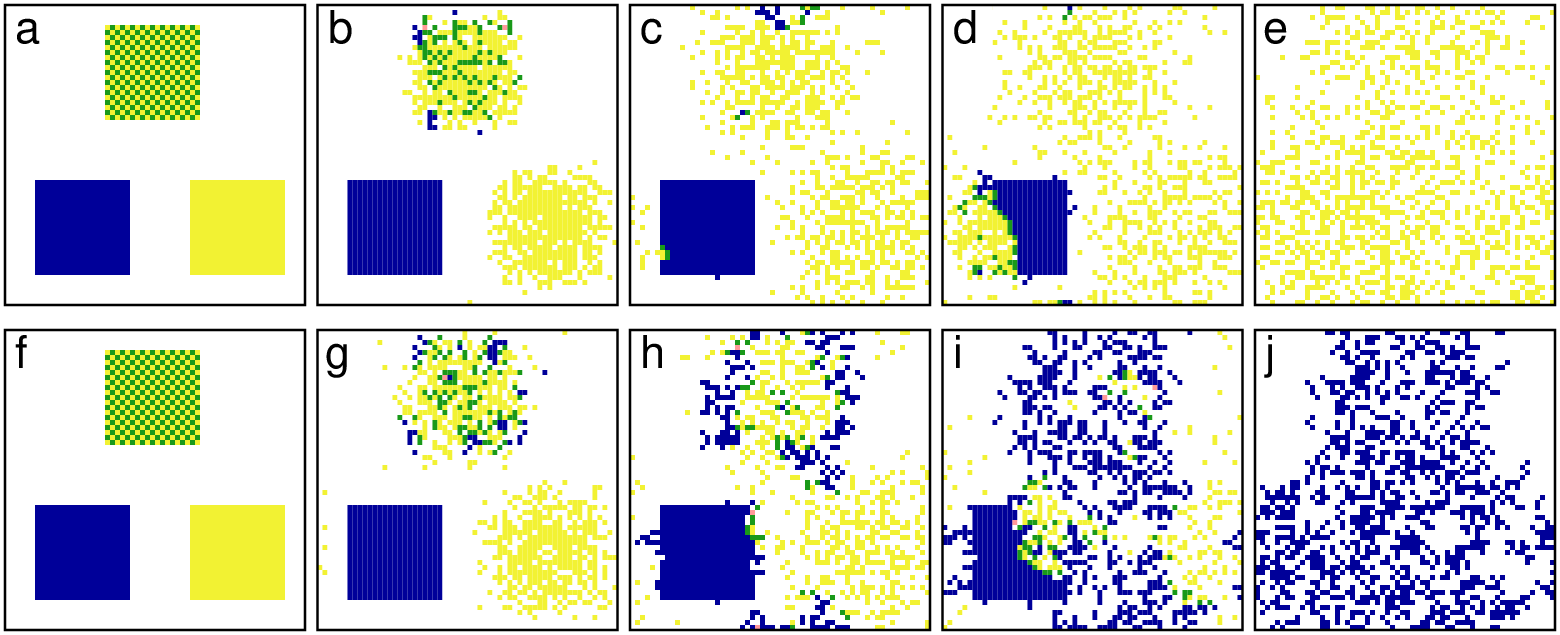}
\caption{(Color online) Pattern formation as observed from a prepared initial state at $\alpha=0.8$. The risk probability parameter is $\beta=0.6$ [upper row from (a) to (e)] and $\beta=0.1$ [bottom row from (f) to (j)]. The color code is the same as used in Fig.~\ref{fig3}, i.e., cooperators whose focal group succeeds (fails) to reach the collective target are denoted blue (green). Similarly, defectors whose focal group succeeds (fails) are denoted pink (yellow). White are empty sites. Other parameters are: $\rho=0.30$, $c/b=0.5$ and $L=60$. Monte Carlo steps are $t=0, 3, 21, 40, 120$ for panels (a) to (e), and $t=0, 8, 40, 85, 380$ for panels (f) to (j), respectively.}
\label{fig4}
\end{figure*}

Figure~\ref{fig3} shows the evolution at the same $\beta=0.2$ value but two different $\alpha$ values. At the low $\alpha=0.2$ value [panels from (a) to (e)], the rare presence of cooperators is already able to ensure the group to avoid the risk. As Fig.~\ref{fig3}(a) shows, blue and pink colors are present in the mixed domains, signaling that both strategies are in successful groups here. As a results, defectors can exploit their advantage by keeping their endowments and become a more successful strategy. The success of defectors will easily drive the community into the tragedy of the commons, which is indicated by the emergence of yellow color here. Figure~\ref{fig3}(b) demonstrates that only those defectors remain successful who are in the vicinity of cooperators (pink is only present in the neighborhood of blue or green). Meanwhile, the initially full $D$ island decomposes since the failure of their groups triggers an intensive migration among these players. When a $D$ player reaches the bulk of the cooperative island, as shown in Fig.~\ref{fig3}(c), it becomes successful and can easily invade the cooperative domain. Naturally, the initially more successful defectors become unsuccessful due to the imitation behind the propagation front, as illustrated in Fig.~\ref{fig3}(d). It is also worth mentioning that this front can only move along one direction because cooperators may imitate the more successful (pink) defectors but the reversed pink to blue or yellow to pink transitions are very unlikely. Finally $D$ prevail and distribute homogeneously due to the permanent migration that originates from their dissatisfaction (constant involvement in high-risk groups).

Keeping the same $\beta$ but choosing a higher $\alpha=0.9$ value, a significantly different evolution can be observed, as illustrated in Figs.~\ref{fig3}(f)-(j). Because of the higher threshold, even the identical initial state is interpreted differently, as shown by different colors in the mixed $C+D$ domain in Fig.~\ref{fig3}(f). Here, the rare presence of cooperators cannot ensure groups to avoid the risk, and hence both strategies are initially stuck in ``unsuccessful'' groups, as evidence by the application of green and yellow color. Because of the high risk both strategies gain nothing, and hence they are neutral. The initial coarsening thus proceeds according to the voter-model-type dynamics \cite{dornic_prl01}. Due to domain growth, when the local density of cooperators reaches the necessary critical level as determined by the $\alpha$ value, cooperators become successful [turning blue in Fig.~\ref{fig3}(g)], which yields a higher payoff for them. Consequently, they start invading the mixed domain, as shown in Fig.~\ref{fig3}(h), because their diluted distribution provides a robust and effective formation against defection. It is important to note that cooperators can become ``successful'' first at the edge of the mixed domain, where the absence of defectors allows them to change their status more easily. This observation helps to understand why the diluted population of cooperators (rather than a compact cluster) is capable to elevate the cooperation level, which will elaborate on later. As can be observed from in Fig.~\ref{fig3}(i), the above mentioned diluted distribution of cooperators can invade the full $D$ domains, ultimately resulting in the total victory of cooperation. The final state, plotted in Fig.~\ref{fig3}(j), also illustrates that the very high threshold makes the compact $C$ domain robust against the attack of drifting defectors. The latter fail to exploit cooperators, and instead change their strategy upon contact. From this point onwards they stop moving, which practically results in a growth of the cooperative domain, similar to the well-know diffusion-limited aggregation \cite{witten_prb83}.

The influence of $\beta$, which determines the shape of the risk probability function, is illustrated in Fig.~\ref{fig4}, where the same threshold parameter $\alpha=0.8$ was used for both series of snapshots. Here we compare the evolution at $\beta=0.6$ [upper row, from (a) to (e)] and at $\beta=0.1$ [lower row, from (f) to (j)]. For both initial states plotted in panels (a) and (f), the relatively high $\alpha$ value results in ``unsuccessful'' $C$ and $D$ players in the mixed domain. Accordingly, a random drift of both strategies starts the evolutionary process, similarly as described for Fig.~\ref{fig3}(g). The randomly aggregated cooperators can form the seed of a successful ``blue'' domain, but because of the relatively high $\beta$ value the neighboring defectors will not necessarily fails and might keep their endowments. Consequently, cooperators cannot utilize the advantage of the high threshold and loose. As always, the pure defector domain falls into pieces intensively. When a $D$ player reaches the pure $C$ domain, as shown in Fig.~\ref{fig4}(c), it can invade it entirely and fast.

This invasion, however, is significantly different from the one previously described for Fig.~\ref{fig3}(d). In the present case defectors cannot be focal players of a successful group because of the high $\alpha$ but, on the other hand, can avoid the risk (due to large value of $\beta$) and are capable to lower the payoff of ``green'' cooperators who are located at the frontier of the full $C$ domain. Consequently, the weakened $C$ will adopt the $D$ strategy, and hence shift the invasion front further. In contrast to the previously mentioned front propagation, the present movement can be reversed, which depends sensitively on the $\beta$ value. The final state, plotted in Fig.~\ref{fig4}(e), is a randomly distributed full $D$ state.

\begin{figure}
\centering
\includegraphics[width=6cm]{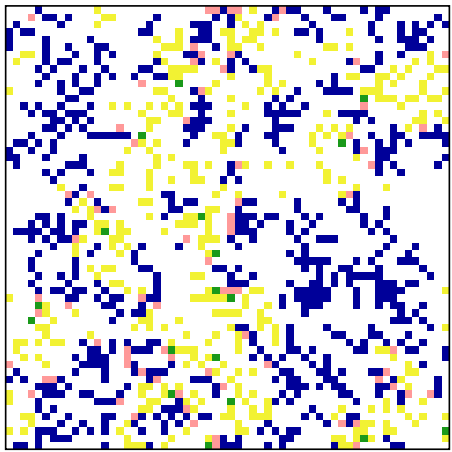}
\caption{(Color online) Representative snapshot of a stationary mixed phase, as obtained for $\alpha=0.4$, $\beta=0.9$, $c/b=0.3$ and $\rho=0.30$. The color code is identical to the one used in Figs.~\ref{fig3} and ~\ref{fig4}.}
\label{fig5}
\end{figure}

If using the same threshold but at lower $\beta$ value, as plotted in Figs.~\ref{fig4}(f) to (j), the outcome is rather different. As panel (g) shows, $C$ players can aggregate randomly again to become ``successful''. Temporarily even $D$ players can manage to be central players of a successful group, yet the low $\beta=0.1$ will result in a high risk probability for defectors who cannot prevail. Instead, as panels (h) and (i) illustrate, the rare distribution of $C$ players prevails again. This effective formation spreads from the edge to the center of the mixed domain as mentioned earlier. Meanwhile, when a drifting defector reaches the compact $C$ domain, as shown in Fig.~\ref{fig4}(h), a different type of invasion can be observed. Since the threshold is not extremely high, a defector who is surrounded by cooperators can collect competitive payoff which is necessary to intrude the pure $C$ phase. Behind the invasion front a mixed phase emerges, as shown in Fig.~\ref{fig4}(i). This mixture of ``unsuccessful'' cooperators and defectors is an excellent target to invade by the rare distribution of cooperators, as we have already described for the initially mixed domain. In other words, compactly clustered cooperators will loose the battle, but the evolutionary process will ultimately be won by diluted cooperators later.

\begin{figure}[b]
\centering
\includegraphics[width=8cm]{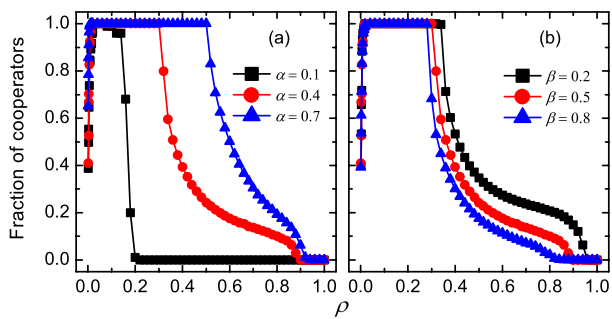}
\caption{(Color online) Fraction of cooperators as a function of the population density $\rho$ for a fixed value of $\beta=0.5$ and three different values of $\alpha$ in (a), and for a fixed value of $\alpha=0.4$ and three different values of $\beta$ in (b).}
\label{fig6}
\end{figure}

In the above discussed cases the final state was always a full $D$ or a full $C$ phase. There are, however, parameter values where the stationary state is a mixture of both strategies. Here, as Fig.~\ref{fig5} illustrates, the diluted cooperators compete with temporarily aggregated defectors. At the edge of these fronts either rare, and hence successful defectors invade, or dense, and therefore unsuccessful defectors are invaded, finally resulting in a dynamical balance of the competing spatial patterns and the stable coexistence of the two strategies.

In what follows, it remains of interest to show how the population density $\rho$ influences the evolution of cooperation. Figure~\ref{fig6} presents the main results to that effect. It can be observed that the fraction of cooperators first sharply increases and then decreases while the population density is increased. As we have discovered recently, low population densities can be extremely useful for the successful evolution of cooperation \cite{wang_z_pre12b}. Here we confirm this, although by means of a self-organizing process, discovering that a full $C$ phase can be obtained for sufficiently small population densities for different values of $\alpha$ [Fig.~\ref{fig6}(a)] and $\beta$ [Fig.~\ref{fig6}(b)], even for small $\alpha=0.1$ or large $\beta=0.8$. In general, the region of full cooperation increases with $\alpha$ and decreases with $\beta$. When the population density approaches one, cooperators cannot survive under such adverse conditions. Obviously, the described supereffective diluted formation of cooperators cannot emerge. Nevertheless, for sufficiently favorable conditions cooperators can defeat defectors by forming compact clusters. In the other extreme, when the population density approaches zero, cooperators simply cannot form sufficiently large domains, either diluted or compact, to resist the invasion of defectors. Accordingly, a sharp rise of defectors is unavoidable, as can be observed in Fig.~\ref{fig6}.

\begin{figure}
\centering
\includegraphics[width=8cm]{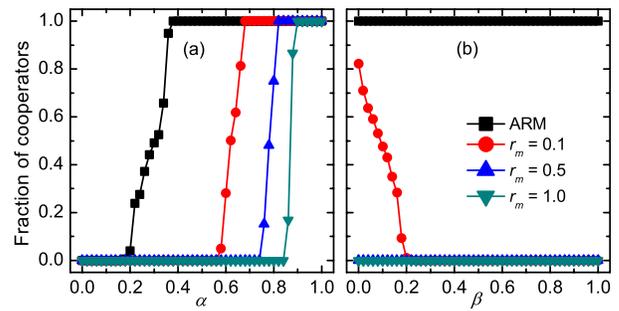}
\caption{(Color online) In panel (a) the fraction of cooperators is presented as a function of
$\alpha$ for a fixed value of $\beta=0.5$ and different constant migration rates. In panel (b) the fraction of cooperators is presented as a function of $\beta$ for a fixed value of $\alpha=0.5$ and different constant migration rates. The population density is $\rho=0.3$ and $c/b=0.3$. Also depicted in both panels are results obtained with adaptive risk-migration (ARM).}
\label{fig7}
\end{figure}

We conclude this section by comparing the efficiency of the proposed risk-driven migration with other game theoretical models, where the collective-risk function is still applied to calculate the payoff but the mobility of players is determined (and adjusted) manually. More precisely, as an alternative model, we apply random migration with a fixed migration rate for each selected individual and compare the outcome with that of the original risk-driven migration model. Figure~\ref{fig7} shows clearly that risk-driven migration always performs better, independently of the $\alpha$ and $\beta$ values. The presented results also evidence clearly that the cooperation level can even be much lower if a higher fixed migration rate is used, which implies that random mobility with a high fixed migration rate is very much detrimental for the evolution of public cooperation. Hence, we conclude that the previous ``migration undermines cooperation'' observation \cite{nowak_11} can only be valid if the mobility is randomly motivated and unrelated with the performance of the group.

\section{Summary}
We have proposed and studied risk-driven migration in the collective-risk social dilemma game on a square lattice. Since the risk was considered to be a dynamically changing group-dependent property that is determined by the difference between the actual contributions and the declared target in each group, the mobility could be introduced in an elegant parameter-free manner, subject only to the self-organization of players according to their experience of risk on different locations. Like previous works considering the impact of mobility on the outcome of games governed by group interactions \cite{wu_t_pre12, cardillo_pre12}, we have discovered that risk-driven migration strongly promotes the evolution of public cooperation. Specifically, the cooperation level increases with the collective target and decreases with the nonlinear parameter of the risk function, whereby complete cooperator dominance was found possible even under extremely adverse conditions. Perhaps even more importantly, we have revealed counterintuitive propagation patterns of cooperative behavior that have thus far not been reported. Quite remarkably, the often cited compact clusters of cooperators have proven inferior to a diluted spatial configuration of cooperators. While defectors could easily infiltrate the former, the diluted population was able to defend itself because upon the invasion of defectors the income in the group dropped suddenly, so that selfishness was unable to grab a hold. What is more, this diluted phalanx of cooperators was able to invade defectors, in turn leading to completely defector-free states. Obviously, the spontaneous formation of this supereffective cooperative formation requires some portion of the lattice to be empty. Accordingly, we have found that there exists an intermediate range of population densities, at which the mechanism works best. This is strongly in agreement with recent results obtained on diluted lattices \cite{wang_z_pre12b}, where the percolation threshold was found close to be optimal for the evolution of public cooperation. Yet there the conditions were given ``manually'', while here they have emerged spontaneously due to risk-driven migration. Moreover, in parameter regions where defectors dominated, we have demonstrated two significantly different invasion possibilities. One under conditions of abundance where groups were in general able to meet the declared targets, and the other under conditions of failure, where the presence of defectors rendered the previously successful cooperators unable to fill the common pool. We have also revealed that the coexistence of cooperators and defectors is possible only in relatively narrow parameter regions, where there is a dynamical equilibrium between the diluted cooperative domains and rare, and therefore successful, defectors and aggregated, and therefore unsuccessful, defectors. Lastly, we have compared the effectiveness of risk-driven migration to random migration with fixed migration rates, and found that for the latter, there exist parameter regions where the evolution of public cooperation is in fact incredibly inhibited. Overall, the self-organizing nature of risk-driven migration outperforms manually adjusted migratory rules by a considerable margin. Our study thus outlines how the dynamical perception of risk can guide migratory patterns in an extremely effective manner, such that public cooperation benefits significantly more than is expected by virtue of the traditional ``compact cooperative cluster'' hypothesis \cite{nowak_n92b}.

\acknowledgments
Support from the Hungarian National Research Fund (grant K-101490) and the Slovenian Research Agency (grant J1-4055) are gratefully acknowledged.

\end{document}